\documentclass[letterpaper]{article}
\usepackage{natbib,alifeconf}

\title{Finding a Mate With Eusocial Skills }
\author{Chris Marriott$^{1}$ \and Jobran Chebib$^2$ \\
\mbox{}\\
$^1$University of Washington, Tacoma, WA, USA 98402 \\
$^2$University of Z\"urich, Z\"urich, Switzerland 8057 \\
dr.chris.marriott@gmail.com}

\begin{document}

\maketitle
\begin{abstract}
Sexual reproductive behavior has a necessary social coordination component as willing and capable partners must both be in the right place at the right time.  It has recently been demonstrated that many social organizations that support sexual reproduction can evolve in the absence of social coordination between agents (e.g. herding, assortative mating, and natal philopatry).  In this paper we explore these results by including social transfer mechanisms to our agents and contrasting their reproductive behavior with a control group without social transfer mechanisms.  We conclude that similar behaviors emerge in our social learning agents as those that emerged in the non-social learning agents.  Social learners were more inclined towards natal philopatry.  Social learners also evolved a culture of eusociality including reproductive division of labor.
\end{abstract}

\section{Introduction}

Sexual reproduction is a social behavior as two able participants must coordinate their behaviors as well as their positions in time and space.  This social coordination problem is solved by sexually reproducing species in many different ways.  

Some of the behaviors that enhance finding and attracting a mate include herding \citep{R87},  philopatry \citep{CL12}, assortative mating \citep{JBK13} and eusociality.  These behaviors can arise through social mechanisms or non-social mechanisms \citep{WH92}.  

For instance, consider witnessing a herd of animals crossing a plane to drink water from the river.  A well known explanation of the herd is that the animals in the herd follow simple social rules of cohesion, alignment and separation \citep{R87,GO94,PEK99}.  These are social rules and social mechanisms because to follow these rules the agents must be aware of each other and make decisions based on information about others' states.  However, a non-social explanation might be that all the animals were getting thirsty in the sun and independently navigated around obstacles to the river.  With this second explanation, the herd is maintained by the mutual response of the individuals in the herd with no need for social awareness or exchange of information.

Prior work (from now on when we reference the prior work we mean the work in \cite{MC15b, MC15a}) showed that herding, philopatry, and assortative mating arose through non-social mechanisms of convergence and common descent.  This work raised a few questions regarding the role that social interaction plays in many of these observed behaviors.  These mating behaviors can be explained by both non-social and social mechanisms.  In many cases the non-social solution is the simpler one, though it is likely that certain instances of these behaviors indeed rely on underlying social mechanisms.  Further, it is not clear how these behaviors vary relative to the nature of the underlying mechanisms.  

We have augmented the non-social agents from the prior work with both individual learning and social learning capabilities.  Our null hypothesis is that the same breeding structures and organizations will be observed as before.  However, we expect that the new adaptive mechanisms will impact these organizations quantitatively and may lead to other organizations.  Our current work can be compared directly to the prior work but because of implementation differences between the models we have also conducted control tests of our own.  These control tests involve the same agents but under conditions in which social learning and/or individual learning is unavailable.

\section{Social Animals}

Animals display different levels of social behavior \citep{M69}.  Social behavior often is centered on reproductive activities and caring for the young \citep{T72}.  Some social animals can extend this social behavior to other activities like hunting, foraging, and grooming \citep{LW88,B94,CC95,N03}.  The highest categorization of social behavior in animals is \emph{eusocial} \citep{C95}.  Ants, bees, termites, and some mole rats are categorized as eusocial \citep{WH05}.  Humans, of course, are also very social and loosely fit under the eusocial definition \citep{NTW10}.

On the opposite side of the spectrum are \emph{non-social} animals.  Non-social animals engage in the minimal amount of social activity required of a sexual reproductive species, which is to mate.  After mating the mother lays her eggs or gives birth to her young and leaves them to fend on their own \citep{S98}.  All parental investment in child rearing comes prior to birth.

To be classified as eusocial, animals must satisfy three conditions.  Eusocial animals share responsibility for caring for their young, have reproductive division of labor, and have multi-generational communal cohabitation (and also philopatry in some definitions \citep{B00}).  This means that parents live with adult children and older generations help to care for their grandchildren and others.  This, in particular, allows for sharing of learned information across generations.  

We can see that humans loosely fit into this definition.  We certainly share responsibility for child rearing, and we certainly have multi-generational communal cohabitation.  However, we do not have reproductive division of labor in the sense we normally think of.  That is, we don't have a single ``queen'' that births us all after breeding with a few privileged, male courtiers.  However, many still like to apply the label eusocial to humans while some prefer to keep humans in a category of their own.  This debate is not critical to our discussion.

\section{Model}

Our simulation consists of agents in a random geometric network of resource sites.  Each day the agents in a population expend energy to move from site to site, forage for resources, and engage in mating, learning and social learning.  Energy in the simulation normally corresponds to the time an agent can spend doing activities during a day but excess stored energy is also used to reproduce.  The resources gathered during a day determines the energy an agent has for activities in the next day.  The net daily energy gain or loss determines whether an agent lives or dies (if the energy is depleted) and whether an agent is capable of reproduction (if stored energy exceeds a threshold).
%
%

\begin{figure}[!t]
\centering
\includegraphics[width = 240pt]{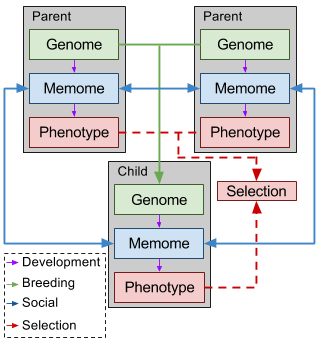}
\caption{Dual Inheritance Model}
\label{fig:dim}
\end{figure}

Agents in our simulation implement the dual inheritance model \citep{MC14, MC16a}. The dual inheritance model is a model incorporating three modes of adaptation: phylogenetic, ontogenetic, and sociogenetic.  That is, agents engage in genetic evolution, individual learning and social leaning.  Figure \ref{fig:dim} shows how genetic and cultural information are stored and transferred in the dual inheritance model.  

Genetic and cultural information are stored in separate information stores called genomes and memomes, respectively.  Genomes are inert within the lifetime of an agent meaning they remain static and are not active in behavior selection.  Their only purpose is to create memomes upon birth and to be replicated in reproductive events.  When an agent is born its memome is created by copying segments of the genome.  This is a process of development that results in the agents initial cultural information.

Memomes are active in behavior selection over the lifetime of an agent and they can also be altered through individual learning and social learning events.  The memome selects behaviors through interaction with the environment and these choices shape the phenotype of the agent.  The phenotype is used for selection and thus an agent's reproductive success and survival is dependent upon its memome, its genome, and its environment.

The cognitive model of our agents is inspired by the pandemonium model \citep{j87,f97}.  The mind of a single agent in our simulation consists of many specialized sub-agents, or daemons, that compete for control of the agent.  In our model we call these sub-agents memeplexes and this internal competition is an evolutionary competition.

\subsection{Genome}

The possible behaviors of agents are encoded in their genomes as a path of resource sites in the network (our model extends the model from \citep{MC15b, MC15a}).  We consider each site in the path a gene in the genome.  
A gene has three components: gathering, non-gathering, and travel.  At each site an agent has an opportunity to gather resources, perform non-gathering actions like breed, learn, and/or socialize, and finally must travel to the next site.  

In our current model an agent always gathers resources when it visits a site.  When not depleted a site will always return a fixed number of resources to an agent gathering at that site.  The energy that an agent spends on gathering is determined by an agent's strategy which is encoded in a gene corresponding to a site.  The energy expended in gathering is always at least the number of resources gathered (one, two or three) and at most five.

A gene for a site will also encode how much time (i.e. energy) is spent performing breeding, learning or socializing actions.  Typically the time performing these actions is considerably less than the time gathering resources.  Actions at a site are performed in the following order: gathering, breeding, learning, socializing, and finally traveling to the next site.  


Agents can engage in sexual reproduction when they have an energy total above a reproduction threshold.  In addition, they need to find another willing and able participant at the same site.  If two agents are engaged in breeding actions at the same site at the same time and they both have sufficient stored energy then they will engage in sexual reproduction.  If no mate is found then agents will wait until the next opportunity to mate.  If the breeding action is unsuccessful an agent must still spend the energy cost of the breeding action but not the cost of reproduction.

A genome is passed to the offspring during reproduction.  In this process, the two parents' genomes are recombined into a single genome and this genome is given the opportunity to mutate.  A genome remains inert during the lifetime of an agent other than during reproductive events.

At birth each agent's genome creates a memome.  A memome consists of a set of memeplexes.  Each memeplex in our model represents a possible set of activities for a day.  Memeplexes are formed by copying segments of a genome.  Starting at each gene (i.e. site) in a genome we copy gene by gene (site by site) into a memeplex (called memes in a memeplex instead of genes).  This continues until the total energy of a segment approaches the maximum energy available to an agent for a single day.  If copying the next gene would exceed the maximum energy the segment is complete and is stored as a memeplex.  This means each memeplex represents a possible set of actions for our agent in a single day.  A memeplex is stored in a memome along with other memeplexes starting with the same initial site for behavior selection (see below).  
 
A single memeplex is formed starting at each gene in a genome.  Copying occurs in the forward direction until the maximum energy is reached.  If the end of a genome is reached copying continues backwards until the maximum energy is reached.  Additionally, segments are copied in a backwards direction from every gene.  This means every gene in a genome is responsible for two memeplexes in its memome except the endpoints that are responsible for only a single memeplex.  Notice that since every site in the environment is not necessarily represented in a genome there may be sites that do not have corresponding memeplexes.

\subsection{Memome}
As mentioned above memeplexes in a memome are arranged by starting site.  The model bears similarities to the MAP-elites strategy of multi-objective evolutionary optimization \citep{mc15}.  For each site in the environment a memome will contain zero or more memeplexes that start at that site.  Each memeplex will consist of a path of gathering sites that begin at a particular site and encode a path that takes at most the maximum energy (time) available to an agent for a single day.  This means each memeplex encodes a possible course of actions for an agent for one day starting at a particular site.  From the perspective of the pandemonium model, each memeplex is a daemon representing a single day's activities. 

At the beginning of each day an agent must select a memeplex that will serve as its behavioral plan for that day.  Only the memeplexes that begin at the current site are possible and so only these memeplexes are considered when selecting behavior.  Since memeplexes in a memome are organized by starting site behavior selection begins by activating all memeplexes that start at an agent's current site.  The agent selects the memeplex from this set that will maximize expected resource gain while minimizing expended energy.  This memeplex is then used to determine the actions of the agent for that day.  These actions interact with the environment to reward the agent with resources which serves as a selective force on the agent.

A memome is not only active in behavior selection but also adaptive over the lifetime of an agent.  It serves as the information storage for both individual and social learning.  This means that additional memeplexes are added to its memome as the agent interacts with its environment and other agents.

If the memeplex determining an agent's actions contains a meme with a non-zero learning component, then the agent will engage in learning during its day.  During individual learning we apply an evolutionary process.  We clone and mutate the memeplex selected for this day's actions and add it to the agent's memome.  

Social learning is handled in the same way as breeding in our agents.  If an agent wants to engage in social learning it must spend time seeking a partner for exchange.  If another agent is also seeking a social learning partner at the same time and at the same site, then the two agents engage in an exchange of memeplexes.  Each agent copies and exchanges the memeplex they used for that day.  In transfer, memeplexs are mutated so noise is added to the system in this step.  The social learning mechanism allows agents to pass their learned memeplexes on to others in the population.  Agents are only allowed a single transfer in a day.  

\subsection{Experimental Setup}
Simulations are run under two conditions: experimental and control.  We ran both conditions 130 times in a variety of different environments.  The experimental simulation consists of a population of agents with all of the mechanisms described above.  The agents in this group are called \emph{socializers}. There were initially two control populations, one in which agents can learn individually but not socially (where agents are called \emph{learners}), and one in which agents can neither learn individually nor socially (and where the agents are called \emph{breeders}).  Learners did not perform significantly differently from breeders in our experiments so our analysis will focus on differences between socializers and breeders.

Each simulation run in our current experiment begins with a population of one hundred agents with randomly generated genomes.  The genes generated at random in this initialization phase have a chance of having a non-zero breeding component to allow for the initial population to breed.  However, all learning and social learning components of randomly generated genes are set to zero.  They can only become non-zero through mutation.
%
%

Simulation runs are seeded with one hundred randomly generated individuals.  Since they are randomly generated any genetic similarity occurs by chance alone.  As shown in prior work it is a rare but fortunate occurrence if two randomly generated agents end up performing breeding actions at the same site at the same time \citep{MC15a}.  Since this is the criteria for breeding in our model this means our initial population has a risk of not being viable.  A run is seeded with one hundred agents so that this chance is diminished.  Under these settings every seed population for a simulation run was viable.

In every run a small proportion of the initial population are fortunate enough to reproduce.  Once initiated these colonies tend to quickly become viable due to forces of common descent.  That is, since the offspring of these reproductive events are related to their parents there is a high chance that they will perform similar behaviors, and importantly, breed at the same time and place as their parents and others in their genetic family.  This helps make the fledgling colony viable.

Each run is allowed to run for 5000 days.  During the run some data was gathered continuously and other data was sampled every 50 days.  For every reproductive event, relatedness of the parents is measured along both genetic and phenotypic lines.  This allows us to monitor the degree of genetic and phenotypic assortative mating present in a run.  Further, whenever an agent dies, the number of offspring an agent had is recorded as well as other characteristics of the agent's breeding history (like breeding sites and breeding partners).

In addition to this continuously gathered data the population is also censused every 50 rounds.  During censusing, the population size is recorded as well as many statistics from every agent alive during that day including: the agent's age, genome length, generation, memeplex generation, as well as the concentration of breeding/learning/socializing components in the agent's genome and memome.

\section{Observations and Discussion}

We indeed saw evidence of herding, assortative mating and philopatry as expected.  In some cases, there are quantitative differences of note.  However, the more exciting results is the emergence of eusociality in our socializers.  Before discussing eusociality let's review some evidence of social learning in our agents.

\subsubsection{Social Learning and Cultural Evolution}

\begin{figure}[!t]
\centering
\includegraphics[width = 240pt]{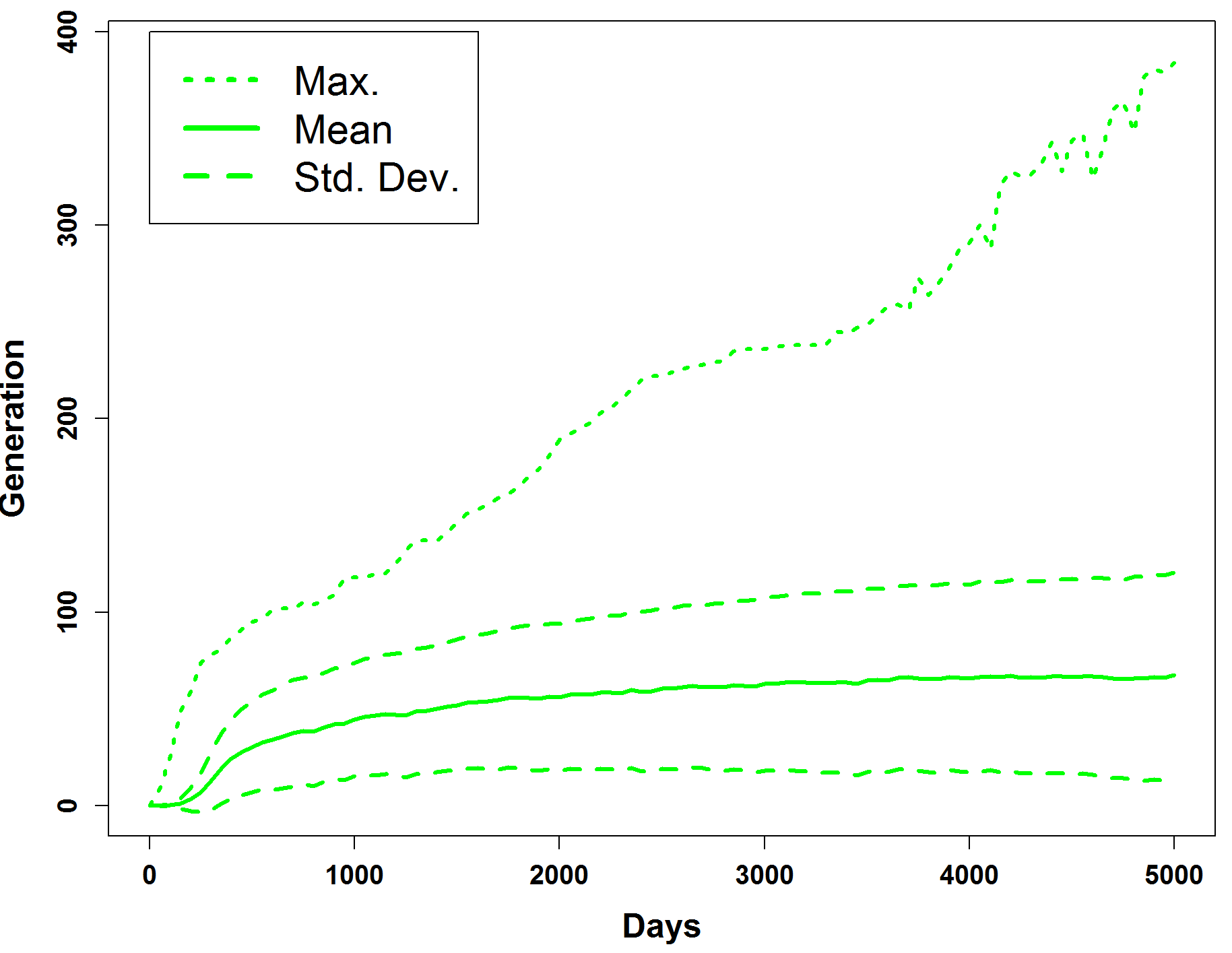}
\caption{Average and maximum generation of active memeplexes over time in socializers with one standard deviation around the mean.}
\label{fig:memegen}
\end{figure}
In order to track social learning and cumulative cultural evolution we assigned each memeplex a generation.  Initial memeplexes created at birth are assigned generation zero.  When a memeplex is cloned the new memeplex has a generation one greater than its parent.  Cloning occurs only during learning and social learning.

Breeders never clone their memeplexes and so they always act on memeplexes of generation zero.  Learners can increase their memeplex generation but cannot share this with others.  Socializers clone their memeplexes in individual learning and in social learning.  In order to test that cumulative cultural evolution occurred in our socializers we have measured memeplex generation over time (see Fig.~\ref{fig:memegen}).

The maximum and average memeplex generation increases over time.  The average memeplex generation grows much slower but the standard deviation also grows over time.  The minimum memeplex is almost always zero because there are usually newborns in the world that have not yet learned the shared memeplex of the population.  A rare cases when this does not occur is during a colony collapse in which there are no newborns (see below).

This is evidence that cumulative cultural evolution is occurring \citep{MC16b}.  This increase over time implies that memeplexes are improved in one generation, shared among the members and passed to the next generation.  Improvements made early are preserved in the population from generation to generation.  We have tracked this optimization over time as well.

\begin{figure}[!t]
\centering
\includegraphics[width = 240pt]{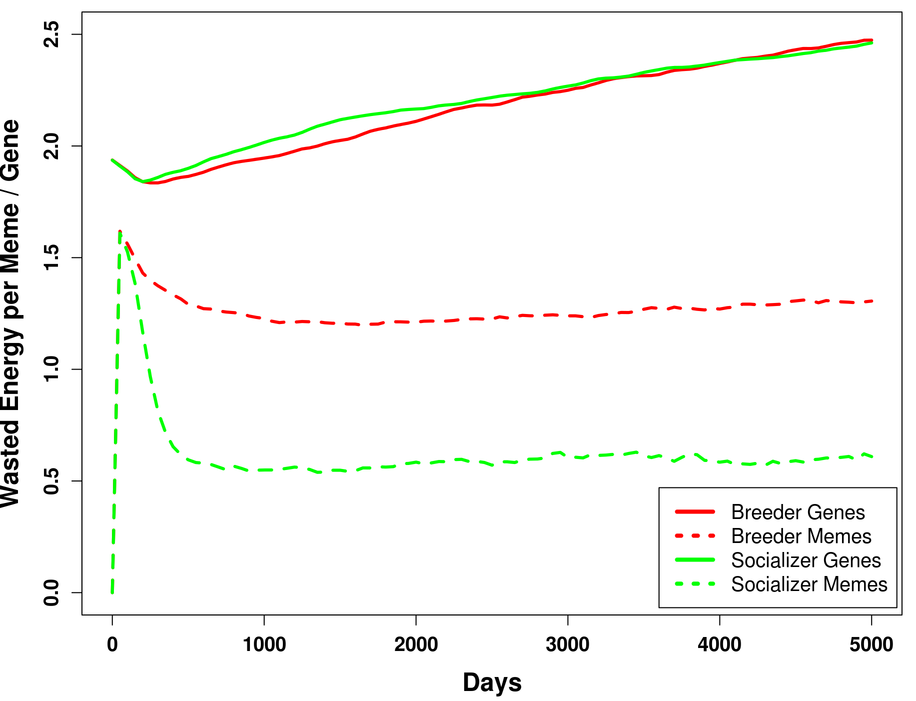}
\caption{Wasted energy in genes in the genome and memes in the selected memeplex are measured.  We show the average wasted energy in the population over time.}
\label{fig:opt}
\end{figure}

In Fig.~\ref{fig:opt} the divergences between gene optimization and meme optimization shows that both breeders and socializers can optimize their behavior.  For breeders this can only occur by optimizing the local parts of their genome that are copied into the memome and eventually used by the agent.  For socializers this means optimizing the shared memeplexes that are passed between agents.  

The data show are averages over the population.  New born socializers have not had a chance to learn the shared memeplexes and thus have memeplexes similar to the breeders that don't optimize.  The most optimized memeplexes in the population have 0 wasted energy after about day 500.  Since this is the most optimized the memeplex can get this slows the cumulative evolution.  The only role of the social learning after this optimization is to maintain the highly optimized memeplex (or one of its clones) from generation to generation.

\begin{figure}[!t]
\centering
\includegraphics[width = 240pt]{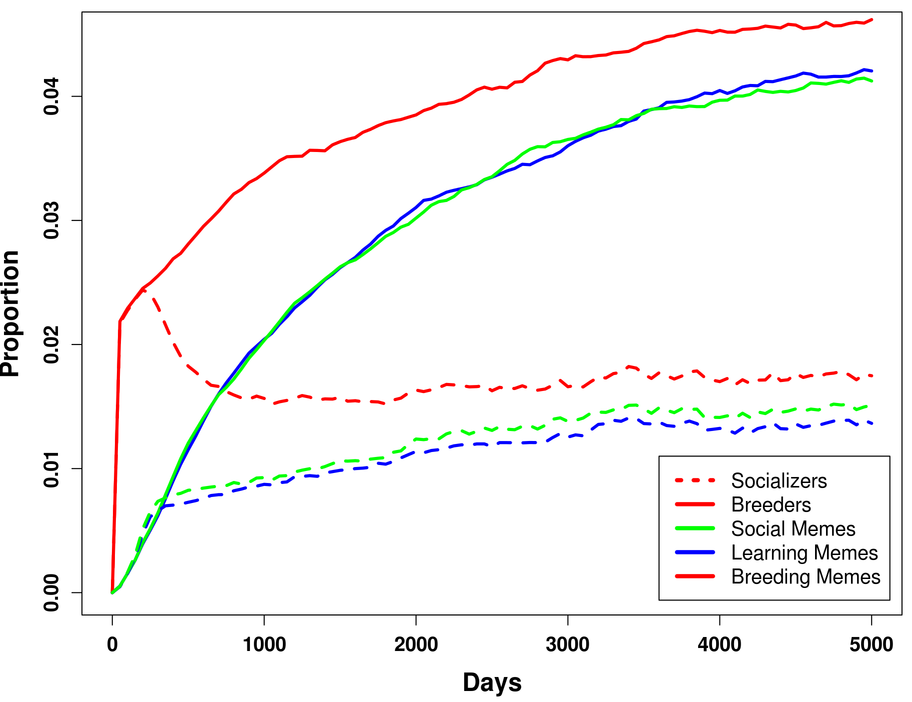}
\caption{Proportion of the selected memeplex devoted to non-gathering actions averaged over all members of the population.}
\label{fig:memes}
\end{figure}

Another means of optimizing memeplexes is to eliminate time spent on the non-gathering activities: breeding, learning, and socializing.  We see (Fig.~\ref{fig:memes}) that breeders evolve to spend more time engaging in these actions over time as selection pressure against this is weak.  Cultural evolution occurs much quicker and so the weak pressure becomes much stronger over the same time.  

We see that the socializers optimize to spend much less time on these actions over the first 1000 days.  The most optimized memeplex would spend no energy breeding, learning or socializing.  This is rare though it occurs.  A memeplex with no breeding component is quite common and suppresses reproduction in the agent if selected repeatedly (see below).  A memeplex with no learning component is not a big detriment since the memeplex is likely already very optimized.  A memeplex without a social learning component makes it impossible to spread itself.  While more optimal these are rare since they die with the host agent.  As a result most optimized memeplexes spend no energy or a very small amount of energy on breeding and usually a little more energy on learning and socializing.  The success of the memeplex depends on its ability to optimize and spread itself.

\subsubsection{Herding} 

Although we have not conducted a quantitative analysis of herding in our agents, we can analyze the underlying mechanisms supporting the herds we observe.  We know from prior work that the herds of the control group are maintained by common descent \citep{MC15a}.  Herds in the socializers are also maintained through the common decent of the memeplexes shared among members of the herd.  Herds observed in socializers, even if outwardly similar, are being maintained by social mechanisms.

\subsubsection{Assortative Mating}


\begin{figure}[!t]
\centering
\includegraphics[width = 240pt]{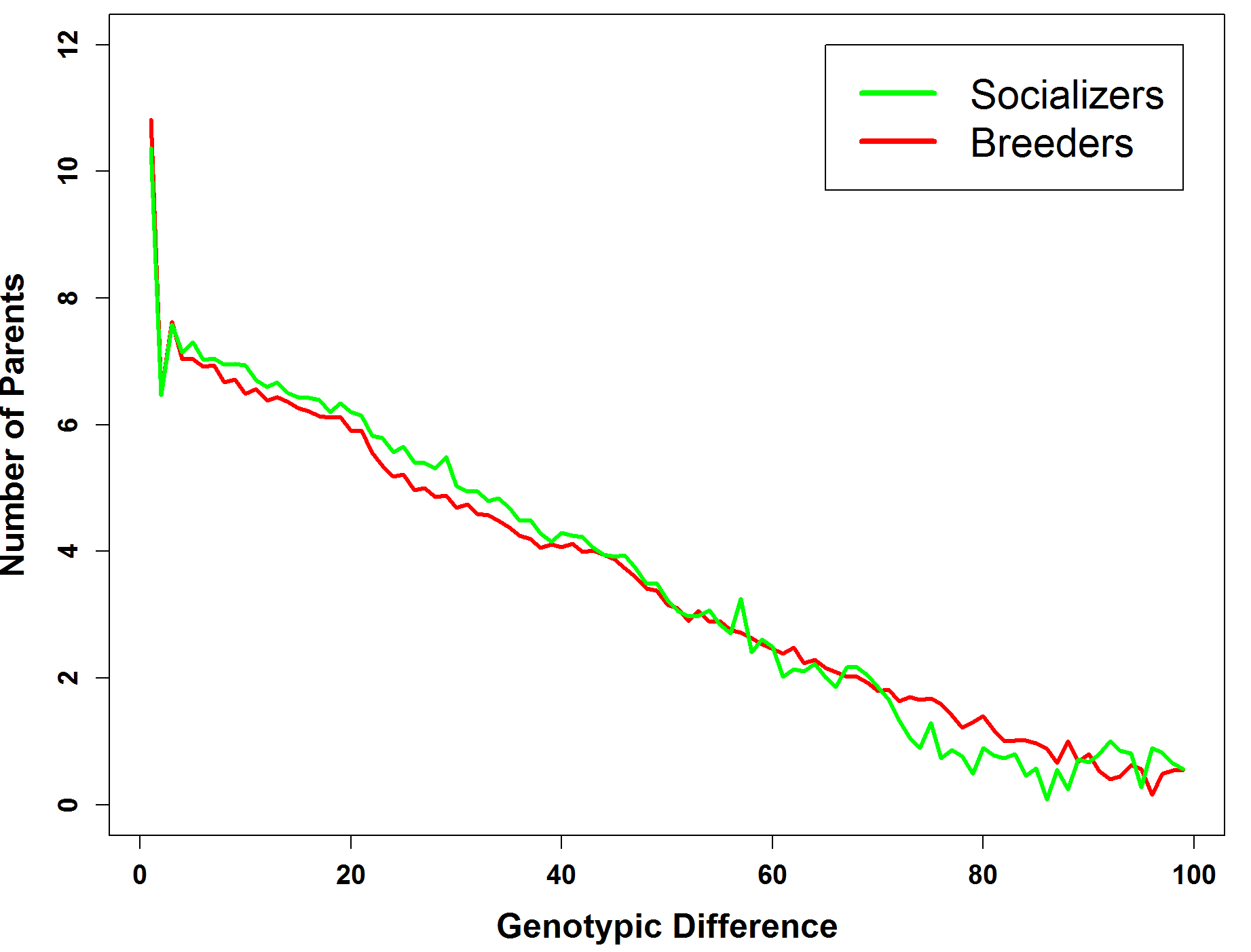}
\caption{Genetic difference between parents.  This data is plotted on a log scale with base 2 and is averaged over the 130 runs.}
\label{fig:geno}
\end{figure}

\begin{figure}[!t]
\centering
\includegraphics[width = 240pt]{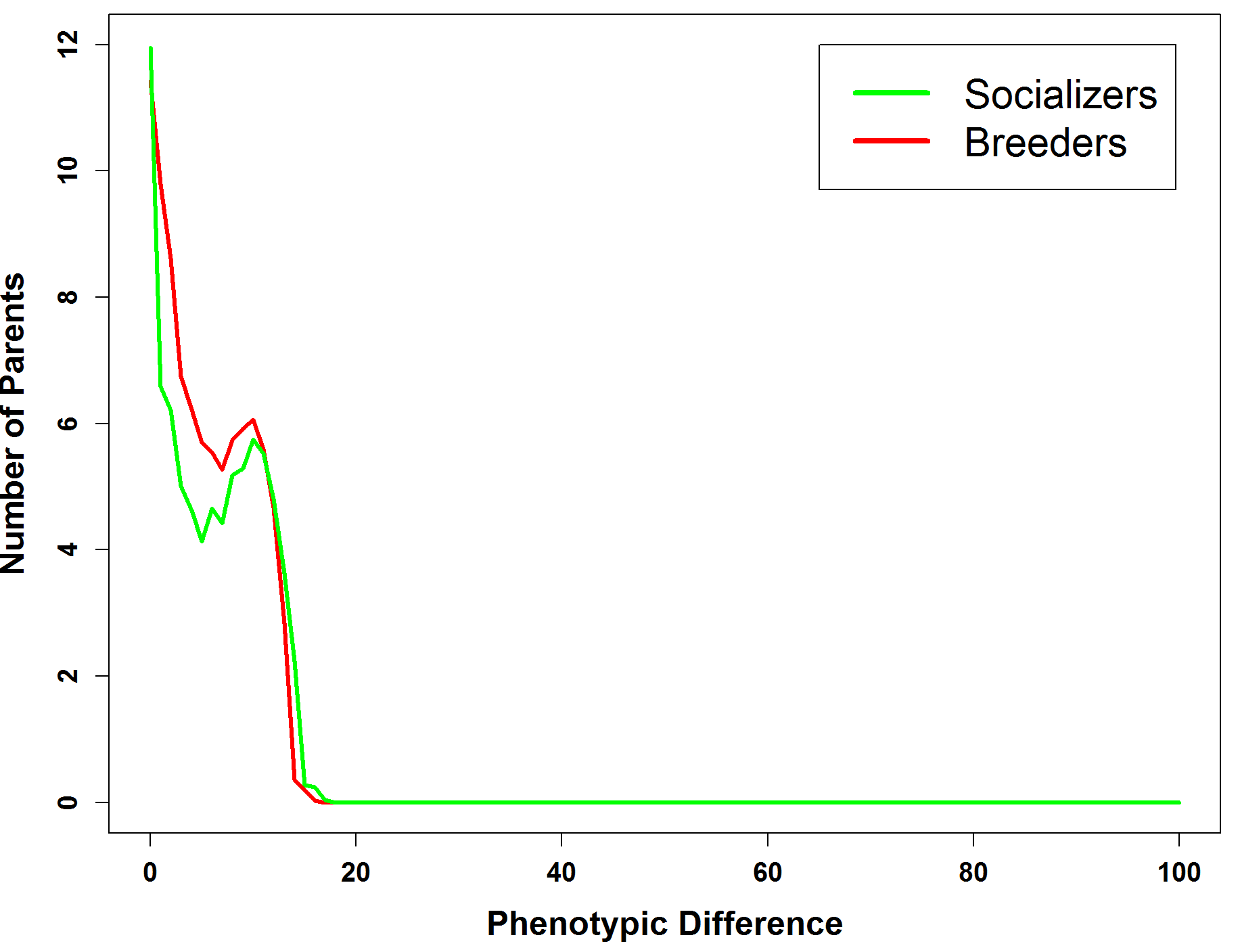}
\caption{Phenotypic difference between parents.  This data is plotted on a log scale with base 2 and is averaged over the 130 runs.}
\label{fig:pheno}
\end{figure}

Assortment of genetically related parents are not significantly affected by the presence of learning and social learning (see Fig.~\ref{fig:geno}).  There are some small differences in assortment of phenotypically related parents though it is not clear how to interpret this slight variation (see Fig.~\ref{fig:pheno}).  



\subsubsection{Philopatry}

When an agent would die we would record it in one of four categories.  If the agent died childless then it could not be evaluated for natal philopatry.  If the agent had children there were three possibilities:  they only bred at their birth site, they sometimes bred at their birth site, or they never bred at their birth site.

Breeders had an average of 45.6\%  childless agents while socializers had an  average of 63.6\% childless agents.  Of the agents that had children,  there were 59.9\%  of breeder agents with children that never bred at their birth site.  In the socializers there were 49.7\% of agents with children that never bred at their birth site.  Fewer agents are breeding in the socializers but more of them are breeding at their birth site.  52.8\%  of breeders that bred at their birth site at least once did not breed elsewhere during their life, whereas in socializers 75.3\% bred only at their birth sites.


While socializers were more likely to die childless they appear to engage in more natal philopatry than the breeders.  They were more likely to breed at their birth site and they were more likely to breed exclusively at their birth site.  

\subsubsection{Eusociality}

Recall the conditions of eusociality among animals.  Agents must have multi-generational communal cohabitation, mutual care for the young, and reproductive division of labor (and sometimes natal philopatry).

Both breeders and socializers have multi-generational communal cohabitation.  We know both display natal philopatry though it is stronger in socializers.  So we must evaluate whether our agents have reproductive division of labor and mutual care for the young.


\begin{figure}[!t]
\centering
\includegraphics[width = 240pt]{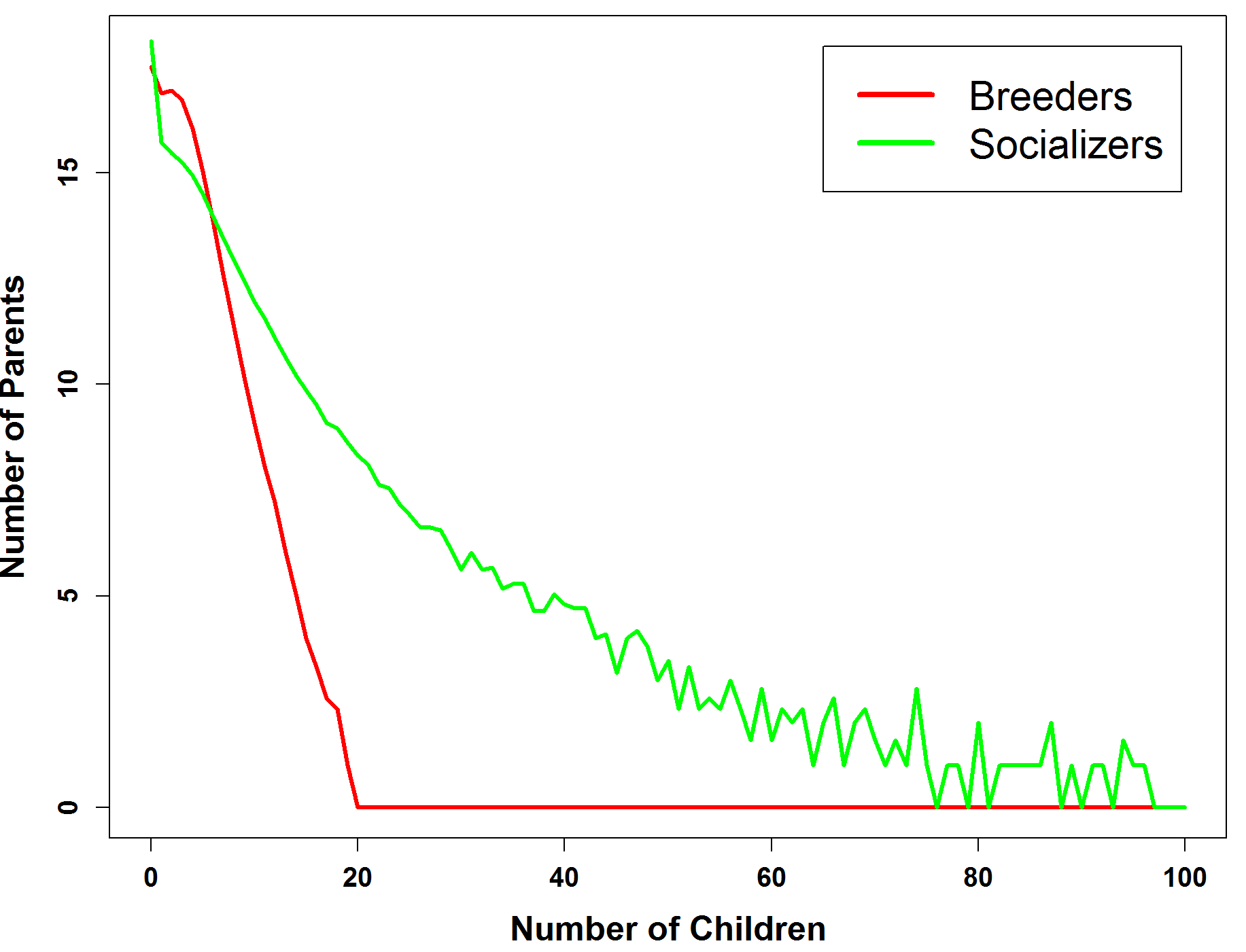}
\caption{The number of children each agent had. This data is plotted on a log scale with base 2.  This data is cumulative over all 130 runs.}
\label{fig:children}
\end{figure}

At death we recorded the number of children the agent had during its life.  We have plotted this data showing how many agents died with $n$ children (see Fig.~\ref{fig:children}).  This plot is on a logarithmic scale and shows an exponential drop off as number of children increases.  Socializers have a  shallower decrease as number of children increase.  No breeders had more than twenty children.  Among socializers some agents, though rare, have more than ninety children.  Combining evidence from above with this we see that both fewer agents are engaging in sexual reproduction and those that do are reproducing more.  This apparently meets the criteria of division of sexual reproductive labor.


We also gathered data on the age of the eldest agent in the population.  We notice that among the socializers there are older agents than among the breeders.
To have many children an agent must live long enough to birth each child and gather the energy required for this activity.  This would require an old agent.  It would also require an optimized memeplex that spent time breeding every day.  If this agent also spread this memeplex to others the efficient breeding culture can be introduced and maintained in the population.

Agents that do not spend energy breeding can save a lot of energy, which can extend their lifespan.  Recall that the most optimized memeplexes spend no time breeding so all energy can be stored for a long productive childless life.  If this agent also spreads this memeplex to others the childless culture can be introduced and maintained in the population.  If the memeplex spread to the whole population then the population will die out (see below).

Among the optimized memeplexes the dominant kinds are those with breeding and those without breeding.  If spread each will generate a different kind of culture.  In our populations these optimized memeplexes occupy agents in the same population.  Not all agents have learned one of these dominant cultures and engage in sub-optimial culture.  This is common of younger agents.  These different cultures compete for the participation of agents.

We treat the sharing of optimal memeplexes to the young as a type of brood care in our simulation as we have not realized a brood care mechanism in our agents.  Using this as a tool to evaluate brood care in our agents we can assess the whether there is mutual care for the young.

Agents in our model do not discriminate when social learning.  They can't recognize their parents or young.  Though there is a higher probability of two related agents occupying the same locations in the environment.  Social learning is also bidirectional.  Both agents act as teachers and learners.  However, usually only the less experienced agent benefits from this exchange.

Now consider an agent with an optimized memeplex that has a breeding component.  After some time this agent produces an offspring.  The newborn has a sub-optimal memeplex and will likely travel a different path than the parent.  If the offspring is lucky in the next few days the agent will encounter its parent and engage in social learning.  Then the offspring may become an optimal breeder as well.  This lucky agent provided direct parent-child brood care.  This is not the only possibility.

The offspring occupies a population with others.  Some may be older siblings, cousins, uncles, etc. while others are more distantly related.  When the offspring first gets a chance to socially learn it might learn from one of these other agents (remember they don't discriminate).  This results in the exchange of sub-optimal memeplexes but possibly also the enculturation of some sub-optimal culture that brings the offspring away from its parent.  We might consider this a kind of mutual brood care.

Finally, let's imagine one of the others in the population has an optimal memeplex that avoids breeding.  This agent cannot reproduce and create its own brood.  It can only spread its memeplexes to the offspring of agents that breed.  Thus, the existence of this culture relies on the care of other agent's offspring.  We have strong evidence that these cultures do indeed exist (see below).

Together this evidence suggests that our socializers have emerged a type of eusociality.  The have multi-generational cohabitation with mutual care for the young.  Like humans they have an interesting culture based division of reproductive labor.  The also engage in natal philopatry more often than the breeders.

\subsubsection{Colony Collapse}
When a highly optimized memeplex has no breeding component but does have social learning it can spread into the population as discussed above.  The danger of this culture is that if every agent in the population follows it then the population will die out.  

This cultural suppression can have catastrophic consequences.  A typical run will begin with a handful small colonies of agents in different parts of the random geometric network.  When a culture of not breeding emerges and spreads to every agent in one of these colonies the population dies out.  In 21 out of 100 socializer runs this led to every agent in the simulation dying before round 5000.  This never occurs in breeder or learner runs.  Inspection of the memeplexes of agents during a collapse confirms that there are no breeding components and most agents have a large store of energy and a long lifespan.

\section{Conclusion}

We were curious how social learning would affect strategies of sexual reproduction in our simulated agents.  We did not see significant differences between the assortative behavior of breeders and socializers.  We interpret this result as additional evidence that assortment can and probably is maintained in most populations by non-social forces. 

Social learning also appeared to enforce a higher rate of natal philopatry.  While fewer social learning agents bred, more of them bred at their birth sites and more of them bred exclusively at their birth site.  This suggests that sociality and natal philopatry may correlate in natural populations.

Finally after adding social learning to our agents we find that they evolve a culture of eusocial reproduction.  Reproductive labor is more concentrated both in a sub-population and occasionally within agents that breed considerably more than others in the population.

All agents that engage in social learning can be considered to engage in brood care by sharing culturally learned information to others.  As they don't discriminate when social learning and there is multi-generational cohabitation this brood care occurs between agents of different generations and of different relatedness.

These are the criteria for eusociality applied to animals and in applying these criteria to our agents we can see there is evidence to call them eusocial.  It is interesting to us that our relatively simple social exchange mechanism is strong enough to evolve a eusocial culture in our agents.  It is noteworthy that this eusociality is maintained by cultural forces not genetic forces.  That is, whether our social agents breed or not is not dependent upon their genetics, but rather on their learned culture.  

\section{Acknowledgments}

The authors would like to thank the advice of anonymous reviewers.  Jobran Chebib was supported by the Swiss National Science Foundation (grant PP00P3\_144846/1 awarded to Fr\'ed\'eric Guillaume).

\bibliographystyle{apalike}
\bibliography{main}  
\end{document}